\documentclass[twocolumn,showpacs,preprintnumbers,amsmath,amssymb]{revtex4}

\usepackage{graphicx}
\usepackage{dcolumn}
\usepackage{bm}

\usepackage{epstopdf}

\begin{document}


\title{Critical heat flux around strongly-heated nanoparticles}

\author{Samy Merabia$^1$}
\email{smerabia@gmail.com}
\author{Pawel  Keblinski$^2$}
\author{Laurent Joly$^1$}

\author{Laurent J. Lewis$^3$}
\author{Jean-Louis Barrat$^1$}

\affiliation{$^1$Universit\'e de Lyon; Univ. Lyon I,  Laboratoire de
Physique de la Mati\`ere Condens\'ee et des Nanostructures; CNRS,
UMR 5586, 43 Bvd. du 11 Nov. 1918, 69622 Villeurbanne Cedex,
France}
\affiliation{$^2$Materials Science and Engineering Department,
Rensselaer Polytechnic Institute, Troy, New York 12180, USA}

\affiliation{$^3$D\'epartement de Physique, Universit\'e de Montr\'eal
Case Postale 6128, Succursale Centre-Ville, Montr\'eal, Qu\'ebec, Canada H3C 3J7}

\date{\today}

\begin{abstract}
We study heat transfer from a heated nanoparticle into surrounding fluid, using molecular dynamics simulations. We show that the
 fluid next to the nanoparticle can be heated well above its boiling point without a phase change.
 Under increasing nanoparticle temperature, the heat flux saturates which is in
  sharp contrast with the case of flat interfaces, where a critical heat flux is observed followed by development of a vapor layer and heat flux drop.
  These differences in heat transfer are explained by the curvature induced pressure close to the nanoparticle,
  which inhibits boiling. When the nanoparticle temperature is much larger than the critical fluid temperature, a very large temperature gradient develops resulting in close to ambient temperature just radius away from the particle surface
\end{abstract}

\maketitle

Sub-micron scale heat transfer is attracting a growing interest \cite{cahill2003}, motivated
by both fundamental and technological issues. The fast emergence of this field is, to a large extent, associated
with the development of micro and nano technologies. In some cases, thermal transfer
 is part of the system function (e.g. the use of nanofluids for heat transport or of multilayered
 materials for thermal insulation). In other cases, the enhancement of heat transfer is a
 key to a proper operation of the microsystem (e.g. microprocessors) and involves
  the integration on ever smaller scales of devices such as micro heat pipes. Although
   these systems are of micrometer size, the regions that limit heat transfer -interfaces,
   constrictions - are often characterized by even smaller lengths, bringing
  heat transfer issues into the domain of nanosciences.
Recent interest in  heat transport around nanoparticles
  has arisen in part from the particular properties of the so called "nanofluids" \cite{keblinski2004,wang2007}, i.e.  colloidal suspensions of solid nanoparticles, which exhibit improved thermal transport properties.
 On the fundamental side, a number of laser heating studies were performed demonstrating even melting of metal nanoparticles without macroscopic boiling of the embedding liquid.~\cite{hartland2004,plech2004}. The physics of this phenomenon involves a complex interplay between boiling, heat transfer, and
particle-fluid interactions (wetting), and is still poorly understood.
\newline
In this letter, we use molecular dynamics simulation (MD) to study heat transfer around a nanoparticle surrounded by a volatile fluid. We characterize
heat transfer in situations where the nanoparticle is heated  above the fluid boiling point and/or critical temperature, and compare this situation to the case of  flat interfaces. We show that  the fluid around nanoparticle can sustain large heat fluxes, well above the critical heat flux of the bare fluid on a flat interface. Accompanying the large heat flux are extreme temperature gradients allowing for localization of the hot liquid to volumes comparable with nanoparticle size.  \newline
MD has been already applied successfully to characterize heat
 transfer across a nanoparticle/fluid interface, in the absence of fluid
 phase change~\cite{vladkov2006,vladkov2008,keblinski2006}. This technique has the
  advantage to give local detailed information on heat transfer and on the
   structure of the fluid close to the nanoparticle interface as well. Furthermore,
   the system at hand is perfectly controlled, which allows to pinpoint unambiguously
    the relevant physical mechanisms at work.
\newline
The  model we simulate  consists of a solid nanoparticle made of $555$ atoms
 immersed in a fluid of $23000$ atoms. All atoms interact through a
  Lennard-Jones potential $V_{\alpha \beta}(r)=4 \epsilon ((\sigma/r)^{12}-c_{\alpha \beta} (\sigma/r)^6)$
   where ${\alpha, \beta}$ refers to solid or liquid atoms. The potential has a cut-off radius $2.5 \sigma$ where
  $\sigma$ is the diameter of the atoms. The parameters $\epsilon$ and $\sigma$ are taken to be the same for both phases.
The parameter $c_{\alpha \beta}=1$ if $\alpha=\beta$; $c_{\alpha \beta}=c$
  otherwise where $c$ controls the wetting interaction between the fluid and the solid nanoparticle. In this work,
  we shall consider the three cases:~$c=0.5,1$, which correspond respectively to
   solvophobic and solvophilic interactions, and $c=2$, which describes strong
   bonding between the solid and the fluid. The solid nanoparticle is obtained
   from a spherical cut of an equilibrium FCC lattice. In addition to the Lennard-Jones
 interactions,  atoms inside the  particles are connected to their neighbors with
  FENE springs~\cite{vladkov2006} $V(r)=-0.5 k R_0^2 \ln \left( 1-(r/R_0)^2 \right)$ with $k=30 \epsilon/\sigma^2$ and
  $R_0=1.5 \sigma$.  This nearest neighbor bonding  allows one to heat up the nanoparticle without observing
  melting or fragmentation.
   This simple  modeling is intended to mimic the situation of metallic or oxide nanoparticles, with rather high cohesive energies,
   in volatile  organic solvents with  much lower cohesion and a low boiling temperature.

\begin{figure}[ht]
\centering
\includegraphics[width=9cm]{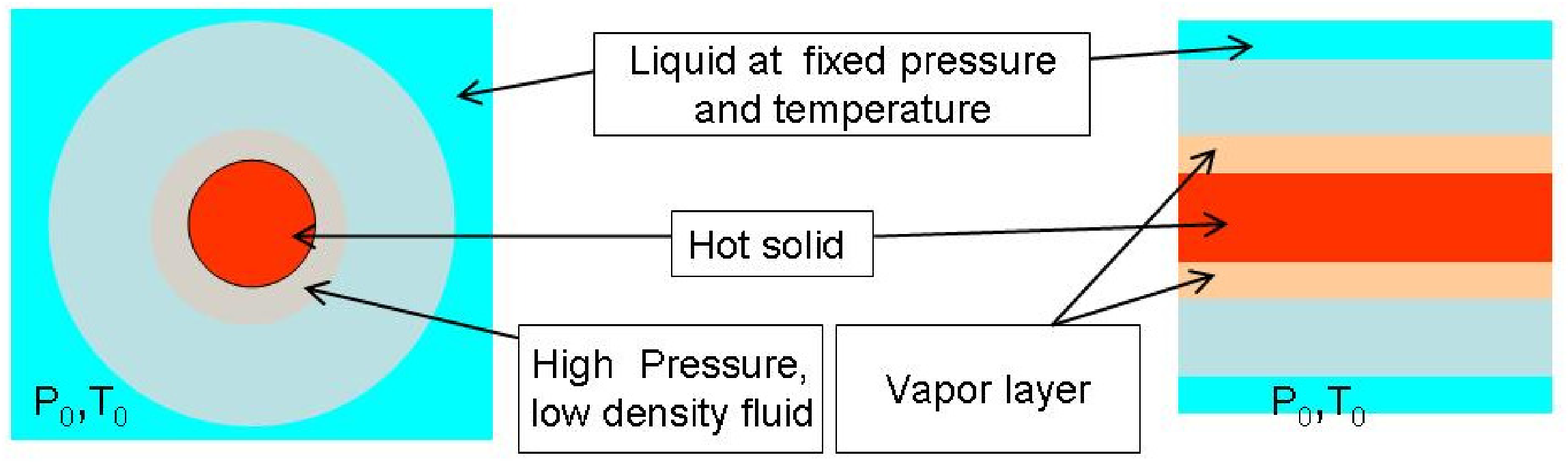}
\caption{\label{setup}Schematic illustration of the simulation setup for the nanoparticle
and planar cases}
\end{figure}

   Throughout lengths, energies and times are expressed in units of $\sigma$, $\epsilon$
   and $\tau=\sqrt{m \sigma^2/ \epsilon}$ where $m$ denotes the common mass of the atoms. For liquid Argon, the
corresponding values are: $\sigma=0.3$ nm, $\epsilon=0.025$ eV and $\tau \simeq 1$ ps.
We integrate the equations of motion using a velocity Verlet algorithm with a time step
$dt=0.005 \tau$. All the systems considered have been first equilibrated at a constant temperature $T_0=0.75$ under the
constant pressure $P_0=0.015$ (using a Nose/Hoover temperature thermostat and pressure
 barostat~\cite{frenkelsmit}). The temperature $T_0$ is below the boiling temperature,
that we found to be $T_b \simeq 0.8$, using independent simulations of a liquid/vapor interface,
under the pressure $P_0$ we are working at. After $100000$ time steps of equilibration, the nanoparticle is heated up
 at different temperatures $T_p>T_b$ by rescaling the velocities of the solid particles
 at each time step, while the whole system is kept at the constant pressure $P_0$
 using a NPH barostat. The fluid beyond a distance $10 \sigma$ from the particle
 surface is thermostatted at $T_0=0.75$, again using velocity rescaling. A global set up of the system is depicted in fig.~\ref{setup}.
Temperature, density and pressure fields have been obtained by averaging the corresponding quantities during
 $10000$ time steps in nanoparticle centered spherical shells of width $\simeq 0.15 \sigma$, after a steady state is reached.
 The Irving-Kirkwood formula~\cite{varnik2000} is used to calculate the normal
 component of the pressure tensor $P_{rr}$. Finally, we calculate the heat flux density
  flowing through the solid particle, by measuring the power supply needed to keep the nanoparticle at the
  target temperature $T_p$.

\begin{figure}[htp]
\begin{center}
\includegraphics[width=6cm]{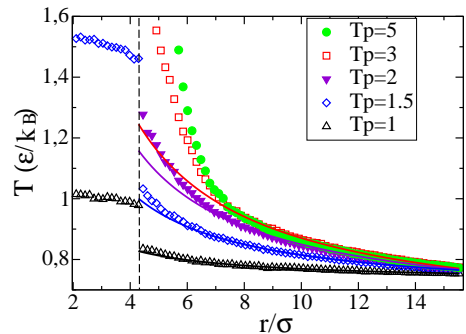}
\caption{\label{temperatureprofile} Steady state temperature field
across the liquid/nanoparticle interface, obtained with molecular dynamics.
The nanoparticle/fluid interaction is hydrophilic ($c=1$). The position
of the nanoparticle surface is indicated by dashed lines, $r$ measuring the distance
to the center of the nanoparticle.
Solid curves correspond to fits by the continuum theory result
$T(r)=A/r+B$. Note that for $T_p=1$, the solid curve is
indistinguishable from the simulation data.}
\end{center}
\end{figure}
Figure~\ref{temperatureprofile} displays steady state
 temperature profiles close to the nanoparticle surface, for different
  temperatures $T_p$ of the nanoparticle. For low $T_p$, the temperature field
   in the liquid is practically indistinguishable from the form $A/r+B$, predicted
   by continuum heat transfer equations in homogenous media in spherical geometry.
Inside the solid, the temperature is not uniform but slightly curved downwards,
due to the finite conductivity of the nanoparticle.
Noticeably, the temperature is not continuous at the solid interface. The corresponding
    temperature jump $\Delta T$ is related to the existence of a finite interfacial
    resistance $R=\Delta T/j$, $j$ being the flux density flowing through the interface.
This effect is well known since the pioneering work of Kapitza on Helium, and is particularly
important when the dimensions of the system considered are comparable to the
 Kapitza length $l_K=\lambda R$, $\lambda$ denoting the thermal
conductivity of the liquid~\cite{barrat2003}. For usual liquids, $l_K$ is on the order of a
few nm \cite{xue2003,ge2006}, thus the effects are particularly important for heat transfer around
nanoparticles. From fig~\ref{temperatureprofile}, we measured a value
 $R \simeq 1.6$, which is consistent with previous MD simulations~\cite{note,vladkov2006} and consistent with
the experimental determinations~\cite{wilson2002,ge2006} in the case of hydrophilic
 interactions, considering that a value of $R=1$ in our units corresponds to
 an interfacial resistance on the order of $0.1$ K$.$m$^2$ GW$^{-1}$. For solvophobic
  interactions, we have measured a larger interfacial resistance ($R=13$ for c=$0.5$) while for
  $c=2$, no temperature jump is seen due to very good thermal contact between the
  nanoparticle and the fluid.  Upon increasing the temperature of the nanoparticle,
deviations from the $1/r$ behaviour are clearly seen on fig.~\ref{temperatureprofile},
when the local temperature exceeds $T \simeq 1$, which is about equal to the critical fluid temperature. Interestingly,
the temperature profile steepens close to the nanoparticle surface,
corresponding to a decrease of the local effective conductivity, and providing "thermal shield" for the fluid away from the particle surface at distances one particle radius. This thermal barrier indicates that the particle can be heated to very large temperatures, while the liquid at particle diameter away from the surface can have close to ambient temperature. Also the temperature gradient at the particle surface can reach enormous values~$0.3$ in reduced units (see Fig. 2) which translates to $300$ K/nm for Argon parameters.
\begin{figure}[htp]
\begin{center}
\includegraphics[width=6cm]{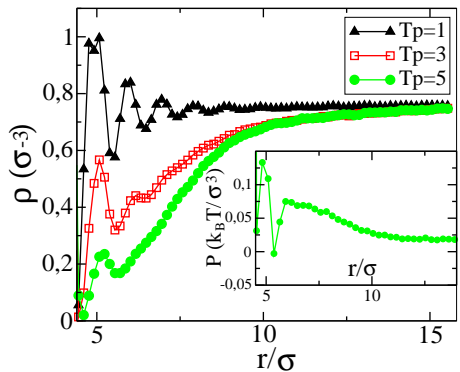}
\caption{\label{densityprofile} Density profile close to a hot nanoparticle having a temperature $T_p$. The inset displays the normal pressure close to the hottest nanoparticle ($T_p=5$).}
\end{center}
\end{figure}
To relate these features to possible structural changes of the fluid,
we have considered the density profile around the nanoparticle in fig.~\ref{densityprofile}.
 When heating is moderate, some layering is present, reminiscent of the layering
 observed at a hydrophilic interface in the absence of heating. When the temperature
 of the particle increases, the height of the first peak decreases while
 layering becomes blurred, and a dilute layer appears.
  Note however that the corresponding density within this layer
   is still one order of magnitude larger than the vapor density at coexistence,
   that we determined using independent simulations of a stable liquid/vapor interface.
   Correspondingly, the inset of figure~\ref{densityprofile} shows a gradual increase of the normal pressure
   in the vicinity of the particle, with values of the pressure approaching the critical pressure,
   estimated to be $P_c= 0.1$~\cite{alejandre1999}. This increase in the local pressure, which is similar
   to a Laplace pressure effect in capillarity,  appears to prevent
    formation of a vapor. It is also worth noting that the decrease of density is accompanied by
   the existence of a limiting temperature profile far away from the particle, as seen in fig.~\ref{temperatureprofile}.
   It is interesting to contrast this situation with the one obtained in a planar geometry.
   To this end, we performed a set of simulations in which a FCC solid slab cut along the $100$ direction
  is in contact on both sides with the fluid. The system is periodic in all directions, and both the solid slab and the liquid
  at a distance $12\sigma$  from the solid are thermostatted  at $T=0.75$.
  The system is kept at constant pressure $P_0=0.015$
   in the direction normal  to the solid slab.
  After equilibration, the temperature  of the solid slab is raised in order to establish a heat flux between the solid
  and the bulk liquid.
  The two situations (nanoparticle and flat solid) are compared in figure \ref{flux}, in terms of the heat flux density
  as a function of the temperature of the solid. In each case, the
  three different wetting conditions ($c=0.5$,  $c=1$, $ c=2$) are considered.
  The difference between the nanoparticle case and that of a flat solid is striking.
\begin{figure}[htb]
\begin{center}
\includegraphics[width=6cm]{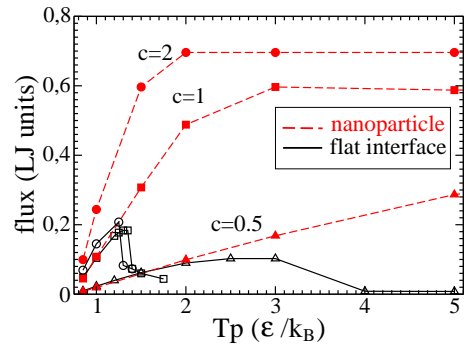}
\caption{\label{flux} Flux density as a function of the temperature of a nanoparticle (filled symbols)
compared to the flux density flowing through a flat interface (open symbols). Lines are guides to the eye.}
\end{center}
\end{figure}
For the nanoparticle, two regimes
can be distinguished, depending on the heating intensity. For low temperatures, the flux
increases linearly with $T_p$, up to a kink temperature $T_k$. The initial slope
 increases as the  wetting becomes better, while the kink temperature $T_k$ moves to higher values with
increasing the solvophobic character of the particle. Beyond  $T_k$ the flux levels off  at a plateau
value which increases
  with the wetting parameter $c$. Interestingly, the transition between these two behaviors takes place when
  deviations from  the $1/r$ profile become significant, and the liquid density of the first peak close
  to the particle decreases significantly. This confirms  that the nature of the fluid in the interfacial zone is
  modified  significantly  for such heat fluxes, with a decreased conductivity - or equivalently a high value of the effective
  interfacial resistance.  For flat interfaces, on the other hand, a quite
   different scenario appears:~for low fluxes, the flux increases nearly linearly with temperature
   and the curves are practically indistinguishable from their nanoparticle counterparts.
   However,  the heat flux density drops abruptly when the temperature is raised above a
   critical temperature that  increases with hydrophobicity. The  corresponding maximum heat
   flux increases with wetting. This  maximum in  the heat flux can be described as a
   "critical heat flux"~\cite{debenedetti} phenomenon, observed here at the atomic scale.
    Note also that this drop occurs below the kink temperature
    $T_k$, but still beyond the boiling temperature $T_b=0.8$.

    From  a microscopic standpoint, the drop in the heat flux is associated
    with the formation of a vapor layer and a nonequilibrium drying of the surface. Due to its low conductivity,
    this layer completely blocks heat transfer from the solid.  The density within the flat vapor layer is  $\simeq 0.02$, one order of
     magnitude smaller than the dilute fluid layer surrounding hot nanoparticles. Furthermore, the pressure within
     this layer is uniform and equal to the imposed pressure $P_0$. This is in contrast with the nanoparticle case,
     where the local curvature of the iso-density curves imposes an increase in the pressure, and in turn
     prevents the formation of a well defined vapor layer. In spite of this difference, a "critical heat flux" can be defined both for the nanoparticle and for the planar walls. Any attempt to
transfer a flux density higher than the limiting values shown  in figure \ref{flux}  would result in an unbounded increase of the temperature of the
nanoparticle, and eventually to its destruction. The critical heat flux, however, is increased by a factor of almost 4 compared to the planar case.
\newline
\begin{figure}[htp]
\begin{center}
\includegraphics[width=5.5cm,angle=-90]{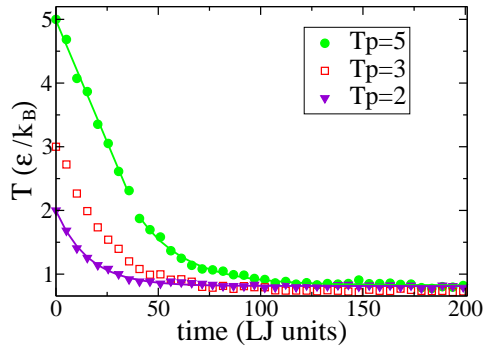}
\caption{\label{relaxation} Cooling kinetics for a wetting nanoparticle ($c=1$), after a steady state was reached, where the particle's temperature was maintained fixed at a temperature $T_p$ .}
\end{center}
\end{figure}
Many experiments that probe  heat transfer at the nanoscale are based on
studies of  time resolved studies of heat transfer in a transient, rather than stationary, regime. We therefore
consider the influence of the critical heat flux on the behaviour observed in such experiments. Our simulation of the transient
regime proceeds by switching off the heating of the nanoparticle and monitoring the cooling process
into the thermostatted fluid.
Fig.~\ref{relaxation} displays the
 cooling kinetics of particles, starting from various initial
  temperatures $T_p$. Two behaviors are to be distinguished. When $T_p$
   is smaller than the kink temperature $T_k$ discussed before (i.e. the initial flux is smaller than critical)
   , the relaxation  is typically  exponential and the interfacial resistance extracted from fitting the relaxation
     curves \cite{vladkov2006} is in good agreement with the value found from the steady state temperature profiles.
     On the other hand, when $T_p$ exceeds $T_k$, the relaxation is no longer exponential
     but initially linear, which corresponds to a constant heat flux flowing through the interface.
     This early relaxation is followed by a second step, where the decay is more exponential,
      with a decay time typically  $20 \%$ smaller than the value reported for lower initial temperature.
       Hence, the change of behavior revealed by our simulations and the existence of a critical heat
       flux around nanoparticles could  be assessed  experimentally
        by monitoring the cooling kinetics in pump-probe experiments. \newline
In summary, we have  investigated heat transfer from nanoparticles and planar interfaces using molecular simulations,
under conditions of high flux and close to the boiling transition of the carrier fluid. In both cases the existence of a critical heat
flux, above which heat transfer is impossible, is observed. This heat flux is considerably higher in the case of nanoparticles, as the formation
of a blocking vapor layer is prevented by the increased pressure around the nanoparticle. This suggests that nanoparticles
could be heated to rather high temperatures inside the host fluid.  The existence of the critical heat flux would manifest
itself by a change of regime in the transient absorption experiments under high flux/high temperature conditions~\cite{hartland2004}. Our study also suggests that nanoscale features could significantly modify heat transfer properties of solid surfaces in this regime. \newline
All the simulations were done using the MD code LAMMPS (http://www.cs.sandria.gov~sjplimp/lammps.html).
This work is supported by the Grant "Opthermal" from Agence Nationale de la Recherche, as well
as grants from the Natural Sciences and Engineering Research Council of
Canada (NSERC) and the \textit{Fonds Qu\'{e}b\'{e}cois de la Recherche sur la
Nature et les Technologies} (FQRNT). We thank the \textit{R\'{e}seau
Qu\'{e}b\'{e}cois de Calcul de Haute Performance} (RQCHP) for computer
resources.


\begin{thebibliography}{99}
\bibitem{note} The interfacial resistance that we measured is larger than the value reported in~\cite{vladkov2006} due to the low pressure
               we are working at.
\end{thebibliography}

\end{document}